\documentclass[preprint]{iucr}              % DO NOT DELETE THIS LINE
\usepackage{miller} 
\usepackage{chemformula}
\usepackage{url}
     \journalcode{J}              % Indicate the journal to which submitted
\begin{document}                  % DO NOT DELETE THIS LINE

\title{Weighted ellipse fitting routine for spotty or incomplete Debye-Scherrer rings on a 2D detector}

\cauthor[a]{Phani S.}{Karamched}{phani.karamched@materials.ox.ac.uk}{address if different from \aff}
\author[a]{Yi}{Xiong}
\author[b]{Chi-Toan}{Nguyen}
\author[c]{David M.}{Collins}
\author[a]{Christopher M.}{Magazzeni}
\author[a]{Angus J.}{Wilkinson}

\aff[a]{Department of Materials, University of Oxford, Parks Road, Oxford – OX1 3PH, \country{United Kingdom}}
\aff[b]{Safran SA, Safran Tech, Department of Materials and Process, Rue des Jeunes Bois – Châteaufort, 78772 Many-les-Hameaux, \country{France}}
\aff[c]{School of Metallurgy and Materials, University of Birmingham, Edgbaston, Birmingham – B15 2TT, \country{United Kingdom}}

\maketitle                        % DO NOT DELETE THIS LINE

\begin{synopsis}
An algorithm for a weighted ellipse fitting routine to measure D-S diffraction ring diameters on 2D detectors in electron or X-ray diffraction.
\end{synopsis}
\begin{abstract}
We introduce a weighted ellipse fitting routine to measure Debye Scherrer rings acquired on 2D area detectors and demonstrate its use in strain determination. The method is relatively robust against incomplete rings due to low number of grains in the diffraction volume (spotty rings), or strong texture (intensity depletion in some azimuths). The method works by applying an annular mask around each diffraction ring and fitting an ellipse, using all pixel positions and their diffracted intensity as weights in the minimisation. We compare this method to the more popular cake integration method, and show that the weighted ellipse method works when the cake integration method fails or works poorly. The lattice strain sensitivity from spotty diffraction rings is in the order or $2 \times 10^{-5}$ or better. The algorithm has been made available for public use and works with 2D diffraction patterns acquired in a laboratory scale XRD equipment, TEM or a synchrotron.
\end{abstract}

\section{Introduction}
%\paragraph[]{\eatpunct}
Microscale deformation is a governing factor for microstructural phenomena like strain accumulation, dislocation motion, crack initiation amongst others that determine how a material behaves under load\cite{Fitzpatrick_2003}. Undemanding laboratory based X-ray diffraction methods for residual strain/stress measurement lack the ability to provide information about the spatial distribution/concentration or have quite low sufficient spatial resolution \cite{WithersBhadeshia}. The most popular techniques at a laboratory scale that provide quite useful information about strain/stress hotspots seem to be Digital Image correlation \cite{MCCORMICK201052} or diffraction based electron microscopy methods \cite{WILKINSON2006307, WILKINSON2012366, BECHE201310}. However, these techniques are limited mostly to surface measurement, and DIC captures strain changes under mechanical/thermal loading rather than initial residual stress levels. High energy synchrotron sources and neutron sources allow these measurements for samples with a much greater thickness, and permit acquisition in a transmission geometry. The nearly parallel beam in a synchrotron also removes the need for a focussing geometry setup and reduced low angle peak broadening effects \cite{Willmott2011}.

%\paragraph[]{\eatpunct}
The framework for stress and strain measurement using synchrotron based X-ray techniques has been quite popular for several decades now \cite{Willmott2011,CullityB.D2014Eoxd}. Incoming X-ray beam entering a material diffracts when the Bragg condition is satisfied.  A 2D or area detector placed orthogonal to the incoming X-ray beam can be used to record the conic projections (Debye Scherrer rings) as whole or portions of rings, depending on the detector size, distance to sample and detector lateral position. The analysis of X-ray intensity of these rings and positions can then be used to determine the lattice plane spacing within the material. With a knowledge of the reference or initial lattice spacing, these spacing measurements can be converted to lattice strain (and through elastic constants, Stress)\cite{ERINOSHO20161,WARWICK20126720}. A geometrically similar situation occurs in the transmission electron microscope (TEM) in the collection of selected area electron diffraction (SAED) patterns.

\begin{figure}
  \includegraphics[width=\linewidth]{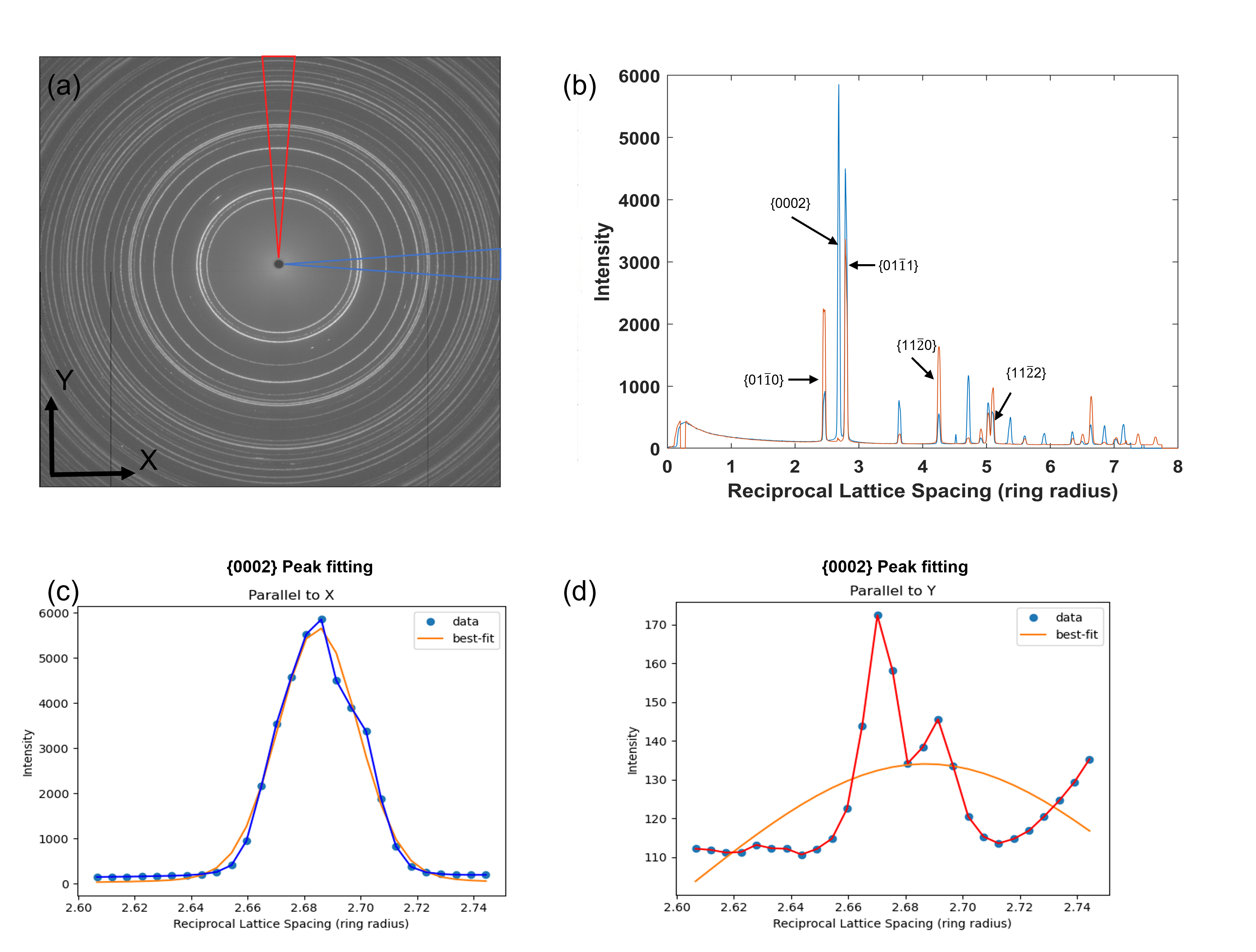}
  \caption{(a) 2D diffraction pattern acquired on an area detector (plotted as a logarithmic intensity), depicting sectors $\pm 5^{\circ}$ around an azimuthal angle $\psi = 0^{\circ}$(i.e. parallel to X in blue) and $\psi = 90^{\circ}$(i.e. parallel to Y in red) from a cp-titanium sample (grade-4), (b) azimuthally integrated in the 2 sectors to show 1D profile of Intensity vs ring radius along X and Y. (c) \& (d) show Pseudo-Voigt peak fitting for \hkl{0002} plane along $\psi = 0^{\circ}$(i.e. parallel to X) and $\psi = 90^{\circ}$(i.e. parallel to X)}
  \label{fig:cp-ti-g4-demo}
\end{figure}

%\paragraph[]{\eatpunct}
As Erinosho \emph{et al.} \cite{ERINOSHO20161} describe in their paper, good quality and representative data can be obtained only when a statistically large number of grains are sampled in the diffracting volume. A statistically large number of grains is satisfied when the individual diffraction spots cannot be distinguished from respective grains and the lattice strain measurement from many grains is representative of the bulk material response. Lattice planes that do not satisfy the Bragg condition remain undetected, leading to a loss of information. Textured materials have an even fewer diffracting grains in some orientations, which often makes the lower multiplicity reflections unsuitable for use \cite{ERINOSHO20161,WARWICK20126720}. Thus, spotty diffraction rings arise when the size distribution of diffracting grains is a large fraction of the incident beam diameter and leads to an orientation distribution lean/skewed. The most common approach to analysing 2D XRD data seems to be adapting a powerful `cake remapping' approach that was introduced by researchers at European Synchrotron (ESRF), built into `FIT2D' \cite{Hammersley1996}, and then re-used in many other software packages that are commonly used to handle and analyse data acquired at a synchrotron source like DAWN \cite{Filik:vg5068}, DIOPTAS \cite{Prescher2015} and pyFAI \cite{Kieffer:gy5006}. This method involves integrating each diffraction pattern along azimutal sectors of defined angular spacing so as to convert them into 1D line profiles of Intensity vs ring radius (or diffraction angle, 2$\theta$) over that azimuthal sector. This integration process requires that the centre of the diffraction ring has already been established with good accuracy. An example 2D diffraction pattern and the resulting 1D intensity profiles generated in two sectors is shown in Figure \ref{fig:cp-ti-g4-demo} (a) and (b).

This approach seems often works reasonably well and the post processing involves measuring shifts in peak positions for all the reflections, which can then be converted to lattice strain with respect to a reference \cite{Willmott2011,CullityB.D2014Eoxd}. The peak position is usually established by identifying an interval for each lattice plane and fitting a curve (usually a pseudo-Voigt function) to determine the peak \cite{ERINOSHO20161}. However, for samples with a strong texture or relatively large grain sizes, some of these sectors may not have any projected intensity of observable reflections or too much scatter in the spots that do not yield a workable 1D profile. This effect of texture is shown in Figure \ref{fig:cp-ti-g4-demo} (c) and (d). This peak corresponds to the \hkl{0002} reflection or the 2nd ring (as measured from the centre) in Figure \ref{fig:cp-ti-g4-demo} (a). The presence of texture in the \hkl{0002} planes does not produce reasonable intensity along the Y-direction and hence results in a poor peak fitting routine, but a good fit along the X-direction. This is also reflected in the intensity of the peak being much lower along the Y direction.

As \emph{Bronfenbrenner} \cite{Bronfenbrenner} shows, any three-dimensional strain manifests as an ellipsoid in the reciprocal space and the projection of this `strain ellipsoid' on a orthogonal 2D area detector takes the shape of an ellipse. Determining the deviation of shape from a circle to an ellipse can be used to extract the change in lattice spacing from a strain-free lattice. This change in the shape of this distorted ellipse with respect to an initial reference can be used to calculate the variation in lattice spacing and lattice strain. In this paper, we explore the possibility, sensitivity and accuracy of fitting an ellipse to these diffraction rings in situations where the `cake remapping' approach may not be appropriate, without a compromise on the data obtained. 

There has been some interest in fitting Debye-Scherrer rings directly in the 2D data as ellipses rather than manipulating the data to 1D intensity profiles.  Most of these cases have focussed on application to detector calibration rather than strain determination.  Hart \emph{et al.} \cite{Hart2013} and Borbely \emph{et al.} \cite{Borbely2014} have developed hybrid 1D-2D procedures in which many 1D profiles spaced around a diffraction ring are used to fit peak positions which are then used as input data for a 2D analysis to fit an ellipse.  Such procedures require assumptions to be made about the geometry so that the 1D intensity profiles are extracted on approximately radial paths.  Hart and Drakopoulos \cite{hart2013weighted} suggest that $\sim$700 profiles be used for continuous rings, but this has to be increased to $\sim$3000 for spotty rings so that sufficient spots are intercepted.  Weighting the importance of the peak positions using their intensity then emphasises the stronger parts of the ring in the 2D ellipse fitting.  We believe the approaches in \cite{Hart2013,hart2013weighted} have been incorporated in the ellipse fitting part of the DAWN \cite{Filik:vg5068} analysis software package. 

A more direct 2D approach was outlined by Hanan \emph{et al.} \cite{hanan2004new}, in which positions of all pixels associated with a diffraction ring were used directly in a least squares fit to an ellipse.  The pixels associated with a particular ring were established by (i) a mask of user defined width around the estimated ring (to avoid overlap with other nearby rings), and (ii) an intensity threshold (rejecting low intensely pixels with the masked region).  This was performed for a series of different intensity thresholds and the intensity weighted averaged used as the best fit for that ring.  

The method introduced here is closer to that by Hanan \emph{et al.} \cite{hanan2004new}, in that it uses a mask to define pixels associated with individual ellipses but then uses a single fit to an ellipse with pixels weighted by their normalised diffraction intensity using an algorithm like that used by Hart and Drakopoulos \cite{hart2013weighted}, in turn a modification of Gal's implementation of a least squares fit to an ellipse \cite{OhadGal}. This method may well be applied for other techniques that involve ellipse fitting to diffraction rings in lab-scale XRD, $\mu$-XRD or TEM based measurement of lattice parameters. The superiority of this method in comparison to the 1D integration routine is shown here for experimentally obtained patterns with incomplete and spotty rings. 

A Matlab routine to perform this analysis has been made available for public use on GitHub \cite{PhaniKaramchedGitHub:Online}. A similar approach in python has also been made available for open source users in the directory. No distortion correction has been applied to the images in this algorithm. The script takes as input the raw diffraction images as \#.\textit{tif} and needs an approximate knowledge of the pattern centre (to apply a mask around individual rings). As suggested by other researchers, a self-convolution of the image can be applied to locate this approximate pattern centre \cite{hart2013weighted,Cervellino:aj5105}. The algorithm estimates a more precise centre of the individual rings, along with the major and minor axes lengths and rotation of the major axis of the weighted ellipse fit. The precise lattice dimensions along the X and Y directions can then be calculated.

\section{Experimental method}
%\paragraph[]{\eatpunct}

Diffraction patterns were obtained as a part of a larger experiment at the I12 beamline at Diamond Light source \cite{Drakopoulos:ie5138}. We first demonstrate the effectiveness of the weighted ellipse fitting routine using a series of ten powder diffraction patterns from a calibration standard \ch{CeO2} (NIST, SRM674b) sample. During the time of recording these patterns, no experimental variation in the sample or geometry is applied. We generate spotty and noisy patterns from this sample and apply the ellipse fitting routine. Noise and spotty pattern have been generated in three ways, (a) by adding white gaussian noise, (b) removing a fixed percentage of points in the pattern at random and (c) removing data along the loading and transverse directions. We compare the cake integration method with a definition of $\pm 5^{\circ}$ sectors and fitting a pseudo-Voigt function to identify the peak position as this seems to be the most popular current approach in the literature \cite{ERINOSHO20161,COLLINS201546} against the weighted ellipse fit method we introduce in this paper. Upon removing data in sectors close to the loading or transverse axis (to simulate the effect of textured samples), the weighted ellipse fit is still able to measure the lattice parameter along the X and Y directions, but the cake integration method fails to work due to absence of intensity data along that direction.

\begin{figure}
  \includegraphics[width=\linewidth]{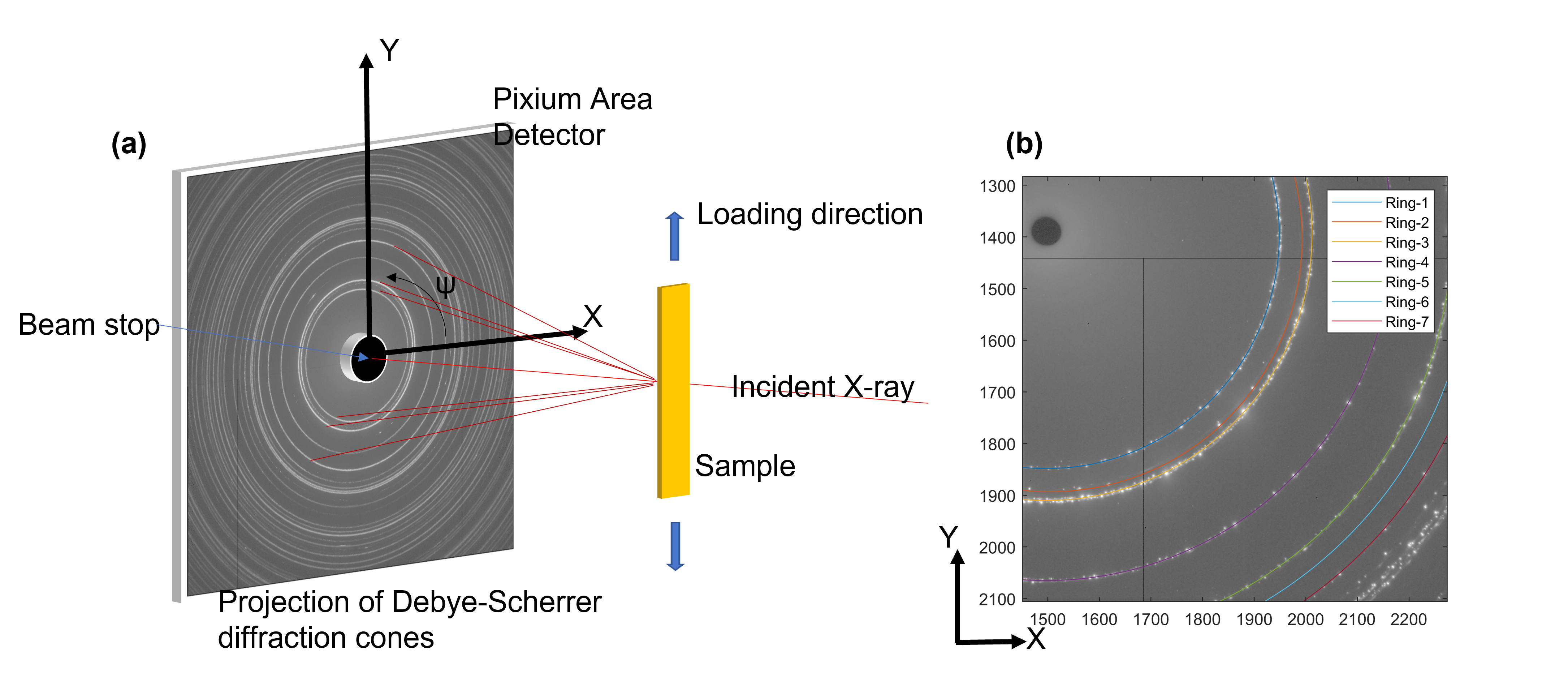}
  \caption{(a) Experimental setup at I12 Diamond Light Source, showing a diffraction pattern from a cp-titanium (grade-4) sample on the Pixium detector (b) Typical spotty diffraction rings (zoomed-in) from a sample of cp-titanium (grade-2) and the working weighted ellipse fit routine.}
  \label{fig:cp-ti-g2}
\end{figure}

As a part of the larger experiment, various Ti samples under were deformed under tensile load in-situ using an Instron Electro-Thermal-Mechanical Tester (ETMT) system (described elsewhere in \cite{XIONG2020561, XIONG2021116937}). A schematic of the setup is shown in Figure \ref{fig:cp-ti-g2}(a). A typical spotty diffraction pattern is also shown in Figure \ref{fig:cp-ti-g2}(b).  Such rings as already discussed are a result of a smaller number of grains/orientations being sampled in the diffraction volume, either due to a larger grain size or texture. An ETMT was placed orthogonal to the incident X-ray beam to perform uniaxial tensile experiments. Tensile load  was applied as shown in Figure \ref{fig:cp-ti-g2}(a) along the Y-direction and the Debye-Scherrer rings recorded on a Thales Pixium RF4343 2D area detector, which has a $2880\times 2881$ X-ray sensitive array of pixels, of size 148 $\mu$m \cite{Drakopoulos:ie5138}.  The image acquisition time per pattern was fixed at 1 s and the incident beam size was set to $1\times 1\ mm^{2}$ on a sample of thickness 1 mm. This weighted ellipse approach has already been used to evaluate the lattice strains for several other reflections in four different cp-titanium grades, the results of which are available elsewhere \cite{XIONG2020561, XIONG2021116937}.

Two sufficiently large grained titanium samples are also compared in this paper. The first dataset is from a cp-titanium (grade-2) sample that has a grainsize of $\sim40\mu m$ and spotty rings (ring-2 and ring-4). The ring-2 and ring-4 correspond to the \hkl{0002} and \hkl{01-12} reflections for the cp-titanium. The second dataset discussed is from a tensile test of the same sample of cp-titanium (grade-2) and the strain measurement from the relaxation part of the test is demonstrated, which normally is a result of cold-creep in these alloys.

The proposed method to fit a weighted ellipse to the diffraction rings uses an approximate ring centre to apply an adjustable annular mask around each ring, and then uses the intensity values as weights to fit an ellipse. This method is almost directly based on the Fitzgibbon approach \cite{Fitzgibbon} and is mathematically simple, easy to apply and computationally very efficient. Analysis of 10 rings in 16-bit depth images on a regular desktop Intel i5 ($8^{th}$ gen) processor and 8gb RAM tends to take of the order of 1 s per image, without parallel computing. The weighted ellipse fitting routine reports an accurate ellipse centre (independent of the input ring centre for the mask), major and minor axis lengths and the semi major axis angle.

\subsection{Ellipse fitting routine}
%\paragraph[]{\eatpunct}
In the cartesian system, the equation for a general conic section follows the form:
\begin{equation} \label{eq:eqn_ellipse}
F(p,x) = p\cdot x = {ax^{2}+bxy+cy^{2}+dx+ey+f = 0} 
\end{equation}
where $p= [a\ b\ c\ d\ e\ f]^T$ and $x= [x^{2}\ xy\ y^{2}\ x\ y\ 1]^T$. The fitting of a general conic can be approached by minimizing the sum of squared algebraic distances of the curve to the points. This methods can be better used by applying appropriate constraints to the conic section and fit an ellipse \cite{Fitzgibbon}. This implementation has already been presented as a remarkedly compact MATLAB implementation by Fitzgibbon \emph{et al.} \cite{Fitzgibbon}, and we modify these equations to make a weighted fit.

The set of points associated with the ring $(x_{i}  ,y_{i})$ are represented as a design matrix defined by $D$ and converted to a weighted matrix $D1$ as shown in equations \ref{eq:matlab_design_matrix}.

\begin{equation}
\begin{gathered}\label{eq:matlab_design_matrix}
D=[x.^\wedge\ 2 ,\ x.*y ,\ y.^\wedge\ 2 ,\ x ,\ y ,\ 1] 
\\
w=repmat(intensities,[1,6])
\\
D1=D.*w
\end{gathered}
\end{equation}

The original set of points are assigned to a design matrix $D$ and converted to a weighted design matrix $D1$, after multiplying with normalized weights as a vector $w$, obtained from the intensities at the positions around the diffraction ring. Mathematical solution follows in line with the Fitzgibbon \emph{et al.} approach \cite{Fitzgibbon} as follows:

\begin{equation}
\begin{gathered}\label{eq:matlab_ellipse_fitting}
\%Construction\ of\ a\ scatter\ matrix\ S  \\
S = D1'*D1\\
\%Build\ a\ 6\times 6\ constraint\ matrix\ C\\
C\ =\ zeros(6);\ C(1,3)\ =\ 2;\ C(3,1)\ =\ 2;\ C(2,2)\ =\ -1;\\
\%Solve\ the\ Eigen\ system\\
[E,V]\ =\ eig(S \backslash  C);\\
\%Find\ the\ positive\ eigenvalue\\
[PosR,PosC]\ =\ max(max(abs(V)));\\
\%Extract\ eigenvector\ corresponding\ to\ positive\ eigenvalue\\
a=E(:,PosC);\\
\end{gathered}
\end{equation}

The original MATLAB based implementation has been preserved in this paper to highlight the difference in the design matrix used from Fitzgibbon's approach \cite{Fitzgibbon}. The logical procedure is described as a comment for each individual command in equation \ref{eq:matlab_ellipse_fitting} as a comment (indicated by a \% symbol at the beginning of each line). The method uses a generalised eigen-system and introducing constraints to specifically fit an ellipse. Amongst the six eigenvalue - eigenvector pairs for this constrained solution, exactly one positive  eigenvalue exists and this unique solution defines the desired elliptical fit. More mathematically trained users are directed to \cite{Fitzgibbon} for further reading.

\section{Results and Discussion}
%\paragraph[]{\eatpunct}

\begin{figure}
  \includegraphics[width=\linewidth]{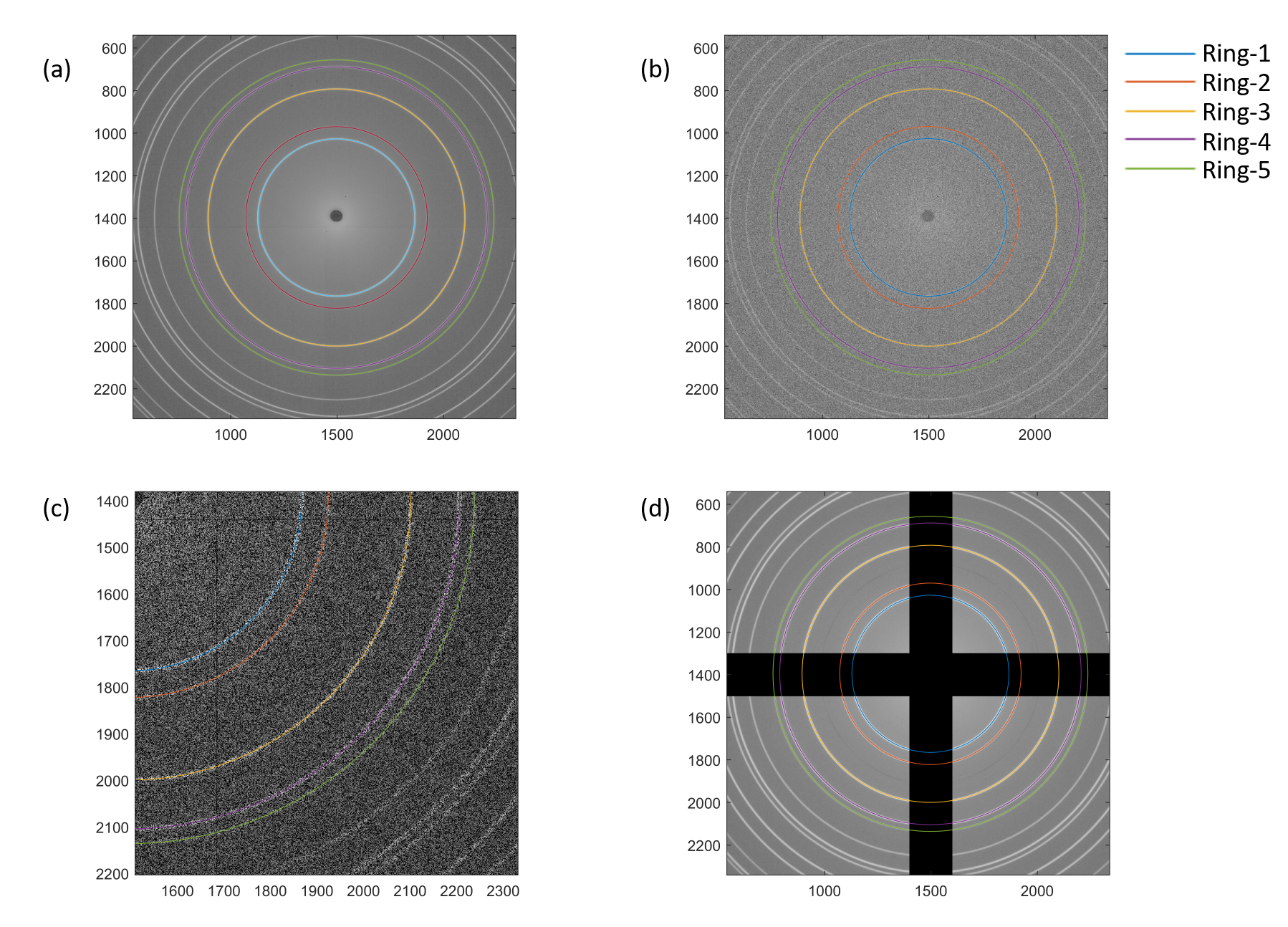}
  \caption{ 5 rings indexed in (a) 2D diffraction pattern from a \ch{CeO2} (NIST, SRM674b) sample (b) the presence of white gaussian noise (c) 50\% of the pixels reduced to zero intensity (d) points removed along horizontal and vertical directions.}
  \label{fig:73014-withrings}
\end{figure}

\begin{figure}
  \includegraphics[width=\linewidth]{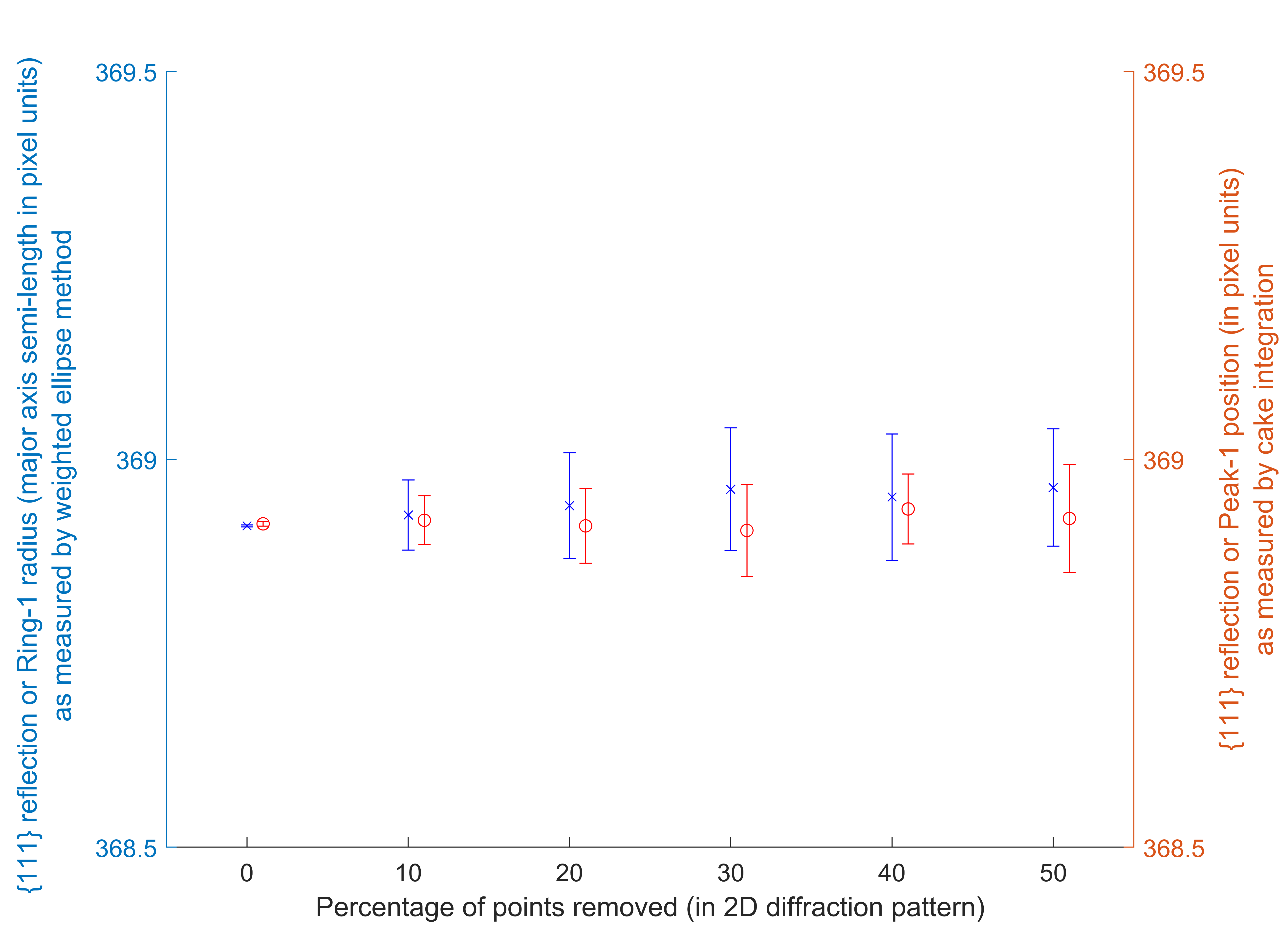}
  \caption{Ellipse major axis semi-length(in blue) or peak positions along Y-direction (in red), measured in pixels via the weighted ellipse fitted method and the cake integration approach for a fixed percentage of randomly zeroed pixels in the image for the \hkl{111} reflection in the \ch{CeO2} sample . The error bars depict the standard deviation in 10 patterns for the first ring/peak. The values on the horizontal axis have been deliberately offset for clarity.}
  \label{fig:73014-hartstyle}
\end{figure}

Figure \ref{fig:73014-withrings} demonstrates that the weighted ellipse fitting routines works even in the presence of noisy patterns from a powdered calibration standard sample of \ch{CeO2}. For raw patterns obtained, the standard deviation in ring-1 (corresponding to the \hkl{111} reflection) semi-length measurement is 0.0013 (in pixel units) via the weighted ellipse fit method and 0.0062 via the cake integration approach. In the presence of white noise, the standard deviation in the peak-1 semi-length measured (in pixels) is 0.3274 using the cake integration method and a pseudo-Voigt fit. The standard deviation measured by the weighted ellipse fitting routine for the corresponding ring-1 is 0.1784. We compare a series of datasets, where randomly a percentage of pixels in the patterns have been brought down to zero. Both the techniques compare extremely well as can be seen in Figure \ref{fig:73014-hartstyle} even when 50\% of the points have been zeroed. This demonstrates that the weighted fitting routine is not only self-consistent, but also consistent with the cake-integration approach. When the intensity along the viscinity of the horizontal and vertical axes are set of zero, the cake integration fails to work, due to the absence of any intensity usable for a pseudo-Voigt fit, but the weighted ellipse fit routine still manages to index as shown in Figure \ref{fig:73014-withrings}(d). The standard deviation measured in the ring-1 semi-length by the ellipse fit is 0.0017 (in pixel units) and closely compares to measurement from the raw patterns.

\begin{figure}
  \includegraphics[width=\linewidth]{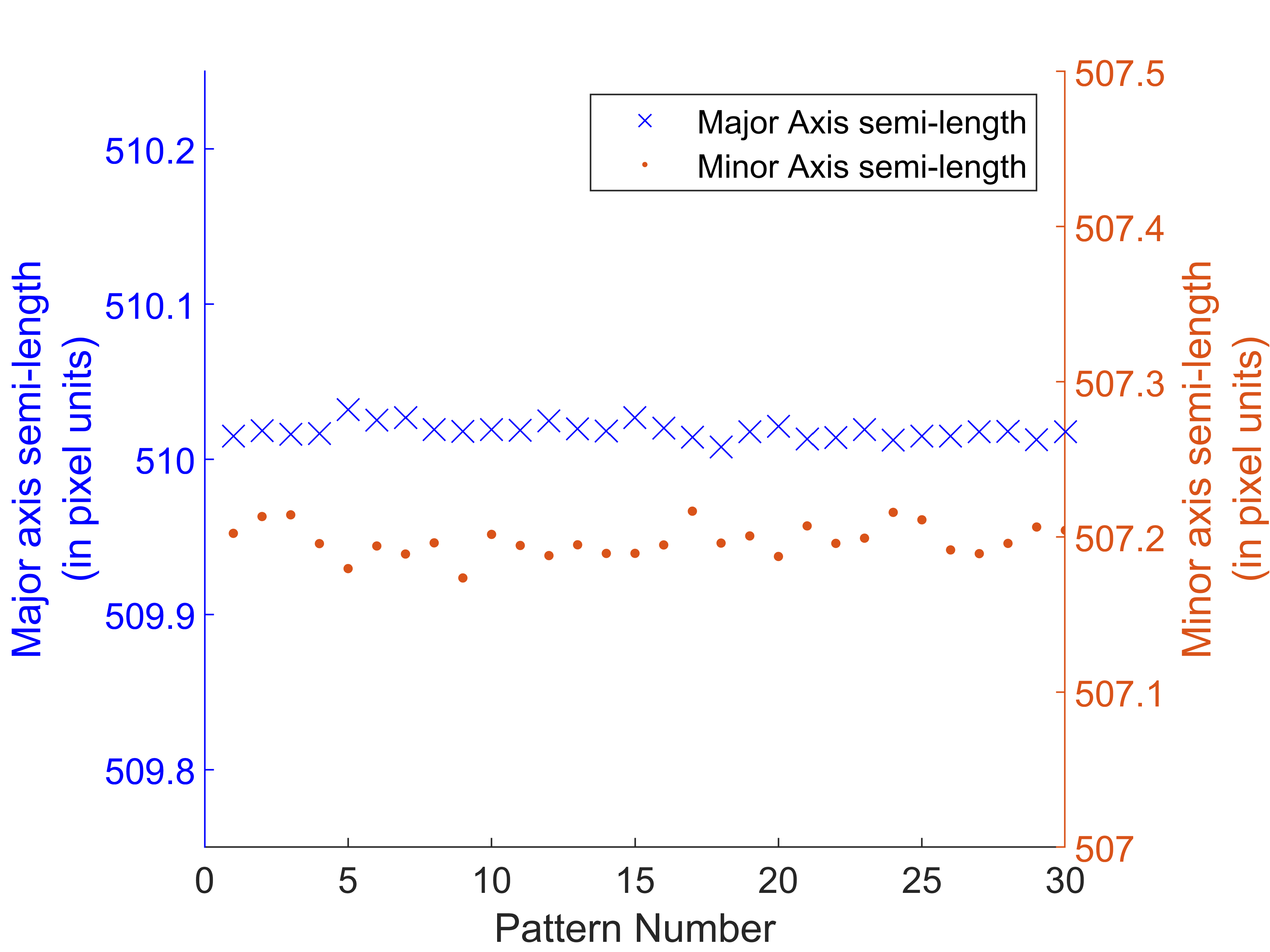}
  \caption{Measured ellipse semi-lengths (ring-2) from 30 patterns in the \hkl{0002} plane of cp-titanium (grade-2).}
  \label{fig:sensitivity}
\end{figure}

%\paragraph[]{\eatpunct}
The two methods are next compared on diffraction patterns from polycrystal (non-powdered) cp-titanium samples (grade-2). Figure \ref{fig:cp-ti-g2}(b) shows the quality of the weighted ellipse fit for spotty diffraction rings and with poor definition of high intensity spots. As with the \ch{CeO2} sample, we measure the deviation on the cp-titanium (grade-2) sample from a recorded set of 30 patterns. During this time sequence of patterns obtained, no experimental parameters  in the sample or geometry were applied. We choose to make this measurement for poorly defined set of intensity on the reciprocal lattice spacing of the \hkl{0002} plane, or ring-2 with sparsely populated high intensity spots. This particular orientation in titanium alloys plays a significant role in fatigue and creep deformation, especially when oriented along the loading direction \cite{Dunne2008_facet,XIONG2020561}. The ellipse fitting routine used to estimate the variation in the ellipse semi-lengths, shows the sensitivity of this measurement from the spotty ring (ring-2) from the \hkl{0002} plane in Figure \ref{fig:sensitivity}. 

The semi-length for the elliptical ring measurement sensitivity of the weighted ellipse technique seems to be better than 0.01 of a pixel. The standard deviation measured in the movement of the centre of the ellipse translated to better than 0.005 in X and 0.007 of a pixel in Y. The standard deviation of the measured lengths along the X and Y axes  of the spotty ring-2 of the cp-titanium sample is $0.0065$ and $0.0075$ (in pixel units) respectively.  This sensitivity in the ellipse determination translates to a strain sensitivity ($\Delta d / d_0$) better than $2\times 10^{-5}$ in these cp-titanium samples, along the direction (Y) with the poorest diffraction information. 

Upon using cake-integration, the 1D intensity profiles generated from the same pattern into 36 {10\textdegree} wide sectors have clearly defined peaks for only 6 of the 36 profiles. Most data points do not have sufficient information to define a peak and follow peak shift measurements. When the data is divided into smaller sectors, the peak definition along the X and Y directions is completely missing or has large errors for lattice strain estimation. This missing definition for the \hkl{0002} reflection (or ring-2) is clearly visible in Figure \ref{fig:cp-ti-g2}(b). With a well-defined ellipse, the strains along any direction orthogonal to the beam direction can be estimated, but without an appropriate peak location, the cake integration method is unusable. The issues with the peak-fitting routine from 1-D profiles for a similar sample (having a better intensity definition) have already been shown in Figure \ref{fig:cp-ti-g4-demo}(c) and (d). 

For a second comparison, we use data from a sample of cp-titanium (grade-2), which has been deformed in an ETMT in the strain relaxed regime. In this comparison we also remove the intensity points along the loading axis of this sample and show that the weighted fit algorithm tracks the ellipse appropriately where the cake integration method is known to fail.

\begin{figure}
  \includegraphics[width=\linewidth]{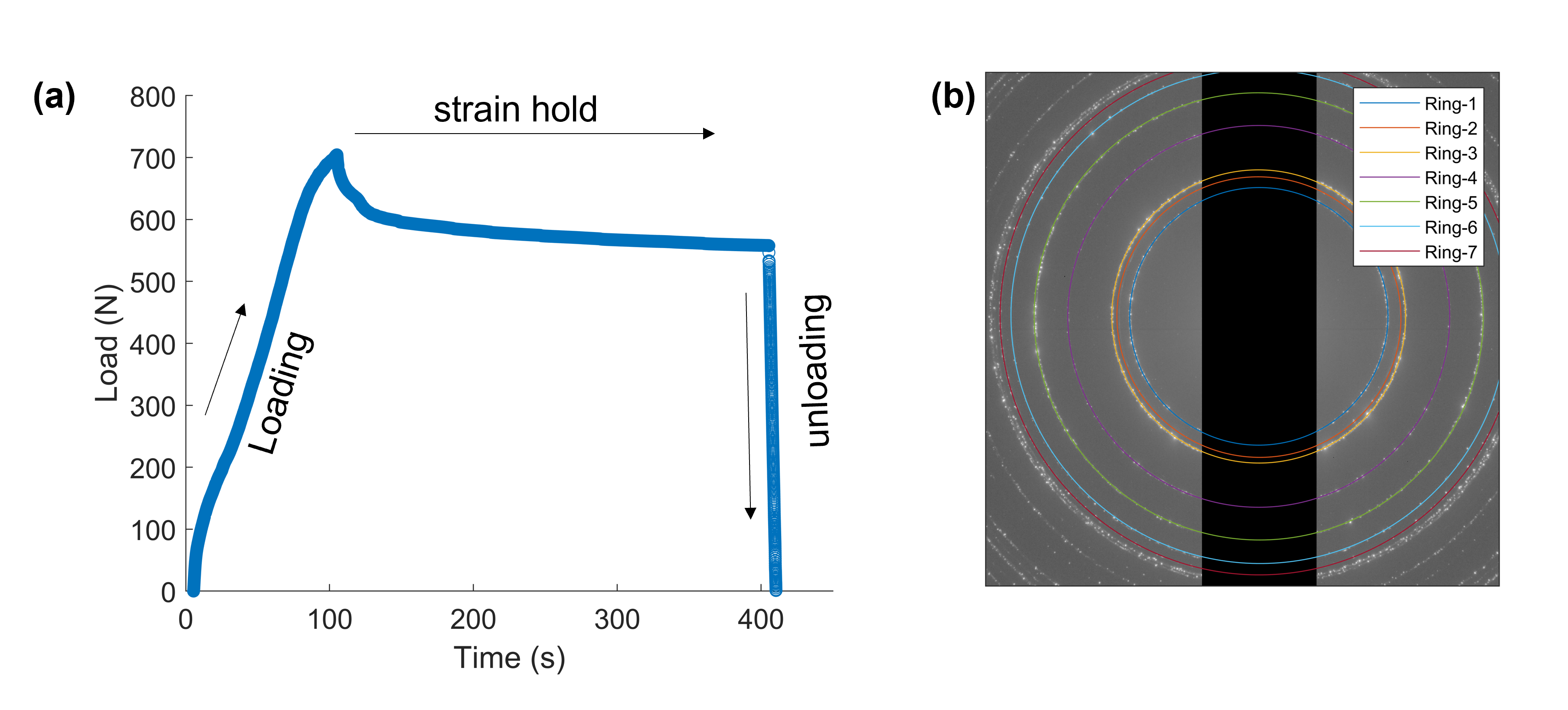}
  \caption{(a) Load (N) vs Time (s) recorded from the ETMT. (b) Mask applied to remove data from the loading direction and weighted ellipse fit.}
  \label{fig:masked+etmt}
\end{figure}

Figure \ref{fig:masked+etmt}(a) shows the loading, relaxation and unloading regime for the cp-titanium sample that was subjected to in the ETMT. After loading slightly beyond the yield point, the stroke arm was held constantly for 5 min. This load (and hence strain) drop during the relaxation period is commonly referred to as cold-creep in these alloys and our primary interest. During this test, the 2D diffraction Debye-Scherrer rings were recorded and lattice spacing measured to estimate lattice strain. We now compare the cake integration vs ellipse fitting routines for ring-2 and also show that despite masking the intensity points along the loading direction, the weighted ellipse fitting routine is able to measure the spacing (and changes) along the Y-direction (loading direction).

\begin{figure}
  \includegraphics[width=\linewidth]{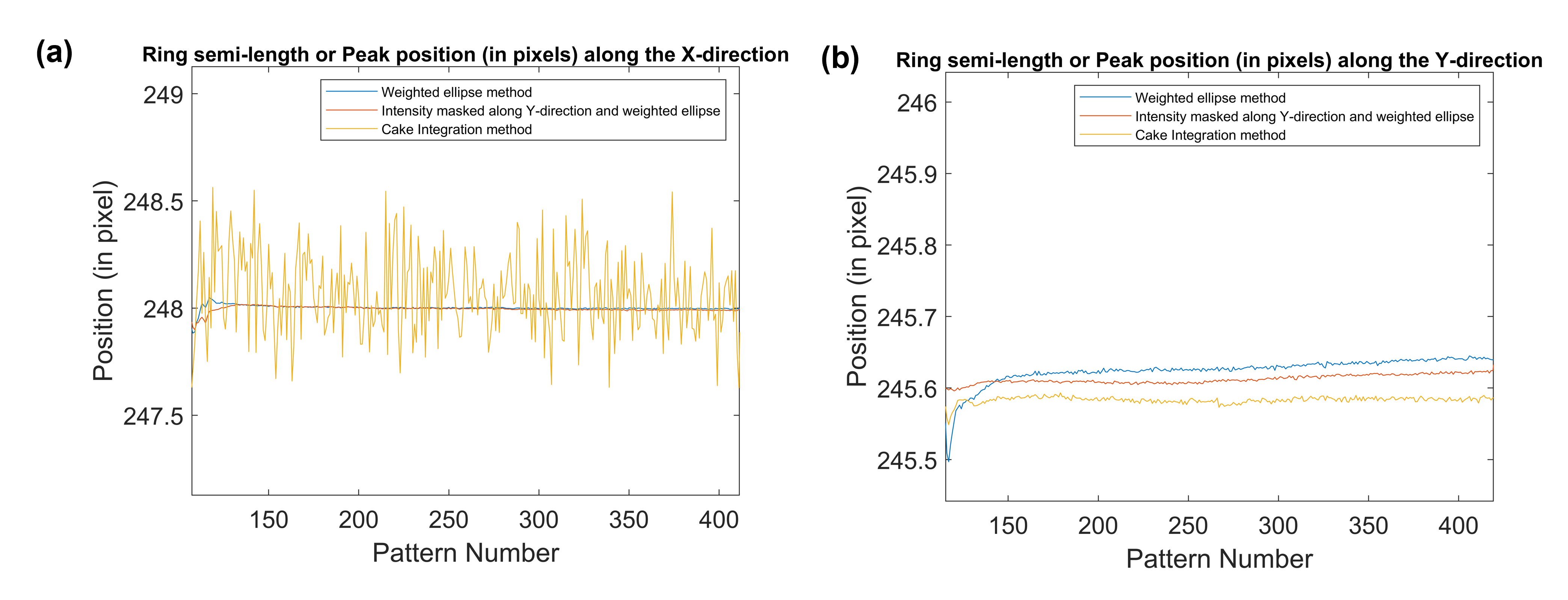}
  \caption{Demonstrating the comparison of the two methods (cake integration vs weighted ellipse fit method) of the reciprocal lattice spacing (in pixel units) of the \hkl{0002} plane along (a) X-direction (b) Y-direction or the loading direction. }
  \label{fig:X+Y-tracking}
\end{figure}

It is expected that the lattice spacing and strains along the X and Y-directions also follow a trend (either tensile or compressive depending on the Poisson's compression and the influence of other crystal planes) as the relaxation shown in Figure \ref{fig:masked+etmt}(a). This relaxation is tracked after a time period of 110 seconds after the start of the test until complete unloading. A comparison is shown in Figure \ref{fig:X+Y-tracking}.

As discussed already, we make this comparison for the measurements of the reciprocal lattice spacing of the \hkl{0002} plane. It is quite clear that in the X-direction, where this particular plane has a poor definition, the sensitivity of the weighted ellipse method is superior to the cake-integration approach. This also holds true when the intensity points are masked along the Y-direction and here the cake-integration approach fails completely. Furthermore, the cake integration method is highly dependent on the calibration used to perform this analysis, that identifies the Beam centre. The weighted ellipse fit is independent of this and a moderately accurate beam centre definition is required only to apply a mask. The centre is then found more precisely from the ellipse fitting. This value of the measured peak position or ring semi-lengths (in pixels) can be converted to lattice spacing (and strain) upon further knowledge of geometry of the setup  and is described in quite a bit of detail in literature \cite{Fitzpatrick_2003, CullityB.D2014Eoxd, COLLINS201546}.

\section{Conclusions}
%\paragraph[]{\eatpunct}
We have thus demonstrated that a weighted ellipse fitting of diffraction rings is a superior, computationally efficient and a more elegant method for identifying/measuring lattice spacing and measuring strain from 2-D diffraction patterns, when compared to the popular cake-integration method. The approach presented here is a combination of several other already existing methods for fitting ellipses and works on raw patterns without a critical need for calibration routines. In cases where the diffraction rings are spotty (due of a smaller number of diffracting crystals in the volume or due to the use of textured samples), the cake integration method may not always be suitable for processing by this method. The weighted ellipse fitting routine presented here and made available for use \cite{PhaniKaramchedGitHub:Online} is more accurate and usually faster.

The weighted ellipse method to measure the ring semi-lengths does not need a calibrated pattern center as is required in the case of the cake integration method. The method only relies on the knowledge of an approximate pattern center to apply an annular ring mask. An accurate pattern center is measured by the fitting routine. Thus, this method can also be applied on a sample regularly used for calibration or for precisely measuring lattice spacing in other methods such as laboratory based XRD or TEM based diffraction. The weighted-ellipse fitting routine also does not assimilate nearby data in comparison to the cake-fitting method that needs a defined azimuthal averaging. The sensitivity in ring semi-lengths measured by the weighted ellipse method is better than $0.01$ (in pixels units) for spotty rings on a cp-titanium sample and the strain sensitivity $2 \times 10^{-5}$ or better. For a standard powder sample of \ch{CeO2}, the standard deviation in the measured value of ring semi-length corresponds to a value of 0.0015 (in pixel units) and in terms of measured d-spacing, this value translates to better than  $1.5 \times 10^{-5}$.

\section{Acknowledgements}
%\paragraph[]{\eatpunct}
The authors acknowledge funding from the EPSRC through the HexMat programme grant (EP/K034332/1) and the Diamond Light Source for beam time under experiment EE17222. We would like to thank Dr.Thomas Connolley, Dr.Robert Atwood and Dr.Stefan Michalik for their friendly and patient help at the beamline I12 and Dr.Nicolo Grilli for the helpful discussion.

\section{Author Contributions}
%\paragraph[]{\eatpunct}
This work formed part of a larger testing campaign proposed by AJW and carried out at Diamond Light Source by YX, PK, CTN, DMC, CMM and AJW. PK oversaw preparation of test pieces and preliminary characterisation undertaken by YX, CMM and PK. CTN and PK setup and guided use of the ETMT system. PK, YX, DMC and AJW developed data analysis procedures. PK wrote the script for pseudo-voigt fitting and ellipse fitting routines. PK drafted the initial manuscript, with all authors contributing to and agreeing the final version. 

\section{List of Figures}

%\section{Supplementary Figures}
%
%\begin{figure}
%  \includegraphics[width=\linewidth]{1D-36profiles.png}
%  \caption{Result of cake integration from cp-titanium (grade-2) sample divided into 36 sectors,
%with {10\textdegree} intervals. Marked in arrow corresponds to the peaks from ring-2.}
%  \label{fig:1D-36profiles}
%\end{figure}

\referencelist[mybibfile]

\end{document}